\documentclass[twocolumn,pra,aps,floatfix]{revtex4}
\usepackage{amsmath}
\usepackage{amsfonts}
\usepackage{amssymb}
\usepackage{graphicx}
\usepackage{hyperref}
\hypersetup{
	colorlinks=true,
	linkcolor=blue,  %
	citecolor=blue,  %
	urlcolor=blue,
	linkbordercolor=0 0 0,
}

\newcommand{\jj}[1]{{\color{red}#1}}

\begin{document}

\title{Non-Markovian Quantum Jump Method for Driven-Dissipative Two-Level Systems}
\author{Huanyuan Zhang and Jiasen Jin}
\email{jsjin@dlut.edu.cn}
\affiliation{School of Physics, Dalian University of Technology, Dalian 116024, China}
\date{\today}

\begin{abstract}
We propose a modified non-Markovian quantum jump method to overcome the obstacle of dramatically increased trajectory number in conventional quantum trajectory simulations. In our method the trajectories are classified into the trajectory classes characterized by the number of quantum jumps. We derive the expression of the existence probability of each trajectory (class), which is essential to construct the density matrix of the open quantum system. This modified method costs less computational resources and is more efficient than the conventional quantum trajectory approach. As applications we investigate the dynamics of spin-1/2 systems subject to Lorentzian reservoirs with considering only the no-jump and one-jump trajectories. The revival of coherence and entanglement induced by the memory effect is observed.
\end{abstract}

\maketitle

\section{Introduction}

Quantum system cannot be completely isolated, it always inevitably interacts with its surrounding environment. The interaction between the system and the environment leads the evolution of the state of quantum system to be non-unitary which is different from the case of a closed quantum system \cite{breuer_book}. To describe the dynamics of open quantum systems, the Markovian approximation is usually employed, which assumes that
the evolution of the system's state depends only on its current state without reference to its history \cite{lindblad1976,gks1976}. In the Markovian dynamics the information of the systems flows unidirectionally from the system to the environment without any feedback \cite{breuer_PRL2009,breuer_RMP2016}. A powerful and efficient approach for simulating the Markovian process is the quantum jump method which considers the stochastic time-evolution of a large number of pure states \cite{plenio_RMP1998,daley2014}. Comparing with the simulation of the full density matrix via the quantum master equation in the Lindblad form, such unraveled process requires less memory since it only manipulates the state vector in each realization \cite{Dalibard1992,Molmer1993}. This has made the Markovian quantum jump method a standard tool in investigating the dynamics of open quantum systems.

However, the Markovian approximation is not always valid for the open quantum systems especially for those strongly couple to the environments or couple to structured reservoirs. In such cases the time-evolution of the state does depend on the history of the evolution by means of, for instance, the information backflow \cite{caruso2014}. These are referred to as non-Markovian processes with memory effects and are found in solid-state physics \cite{lai_PRL2006}, quantum biology \cite{thorwart2009,chin2013,spaventa2022}, and quantum chemistry \cite{shao_JCP2004,pomyalov_JCP2005,chen_SR_2015,PFW2022}. In quantum technologies, the non-Markovianity has been exploited for the entanglement generation \cite{huelga2012}, quantum transport \cite{maier2019,flannigan2022}, quantum metrology \cite{chin_PRL2012,altherr2021,wu2021,bai2021} and the enhancing the performance of quantum batteries \cite{kamin2020,rossini2020,xu_PRE2021,li2022,morrone2023,song2024}. Actually, the non-Markovian process is ubiquitous when the time-scale of resolution is short enough compared to the characteristic time-scale of the system; while the Markovian process is the product of coarse-graining on the non-Markovian process. Therefore, accurately describing and simulating non-Markovian dynamics is crucial for uncovering the underlying physics of these complex phenomena.

The non-Markovian time-evolution of the state of an open quantum system can be described by the master equation with memory kernel \cite{shabani2005,ciccarello2013}. The memory kernel is an integral function about the memory time and weights the past state in the master equation. Besides, the time-local Lindblad master equation can also describe the memory effects in the time-evolution. When the system of interest interacts with a structured reservoir, i.e. the spectral density of varies appreciably with the frequency of the environmental modes, the decay rate becomes time-dependent and even to be negative temporarily \cite{haikka_PRA2010} or permanently \cite{megier_SR2017}. In this work, we will consider the quantum spin interacts with a Lorentzian reservoir. In the limit of weak coupling between the spin and the environment, the time-convolutionless projection technique yields the time-local master equation in Lindblad form describing the time-evolution of the spin under study. In particular, the non-Markovian feature of such master equation is indicated by the dissipator with negative decay rate \cite{haikka_PRA2010,haikka_PhysScr_2010}. The negative decay rate may give rise to the possibility for the system to recover the state that before decoherence. In the framework of Markovian quantum jump method, this can be implemented by performing the reversed quantum jumps to cancel the normal jumps previously occurred when the decay rate was positive. This idea has been put forward in the non-Markovian quantum jump (NMQJ) approach which basically unravels the non-Markovian time-local master equation \cite{piiloPRL2008}.

Despite the significant progress has been made in applying NMQJ to single-qubit system \cite{piiloPRL2008,piiloPRA2009,smirne_PRL2020,chruscinski_quantum2022,donvil_NC2022,becker_PRL2023,settimo_PRA2024}, it is still rather demanding to extend this method to the quantum many-body systems \cite{chiriaco_PRB2023}. The complexity of quantum many-body systems arises not only from internal quantum entanglement and interactions, but also from more intricate interactions between the system and its environment. In general, the system's state after a normal quantum jump is not the eigenstate of the effective non-Hermitian Hamiltonian of the open system and may evolve to excited state that is feasible for next quantum jump.
Therefore the number of the possible path of the time-evolution in many-body NMQJ simulation grows dramatically. Because after a normal quantum jump acting locally on the $n$-th subsystem, the state of the system is a product state of the ground state of the $n$-th subsystem and a state of the rest of the joint system. The normalization coefficient of such state depends on the non-zero amplitude of the state at the moment when the normal quantum jump occurred and thus gives rise to different initial states for the following time-evolution. As a consequence, a large number of time-evolution paths need to be recorded compared with the single-body case. Furthermore, the calculation of the transition probabilities between different trajectories become quite involved and the conservation of the total probability may be broken due to the presence of the reversal quantum jump. Alternatively, introducing an external driving on a single quantum system also presents the challenges similar to those in many-body system, as the number of quantum trajectories likewise grows dramatically. This significant growth not only complicates the probability of reverse jumps in the negative decay rate region but also significantly increases computational complexity, making the NMQJ method difficult to apply in such cases.

In this work we propose a modified NMQJ method that basically dealing with the time-evolution of the existence probability of each quantum trajectory. The existence probability measures directly the portion of a certain quantum trajectory among all the possible trajectories in the limit of infinite number of realizations. We derived the existence probability of trajectory with no quantum jump based on which all the rest existence probabilities can be obtained in a top-bottom manner. This enables us to truncate at an appropriate order in calculating the sum of the expectation value of an observable. We apply the proposed method to the models of a single driven spin-1/2 system and a coupled two-spin system subject to Lorentzian reservoirs. The NMQJ method can effectively simulate the dynamics of the open spin systems especially in the presence of negative decay rate. The recover of coherence and entanglement caused by the memory effects generated by the reversed quantum jump are observed.

This paper is organized as follows. In Sec. \ref{sec_method} we introduce the modified NMQJ method by starting with a brief review of the method of quantum jump in Markovian and Non-Markovian cases. The concepts of quantum trajectory and trajectory class are introduced in Sec. \ref{sec_QTandTC} which are essential in our method. The existence probabilities of each quantum trajectory in the Markovian and non-Markovian cases are derived in Sec. \ref{sec_EP} thus enabling the calculation of the expectation value of the observable. In Sec. \ref{sec_models}, we demonstrate two toy models of spin-1/2 systems that we are going to apply the NMQJ method. In Sec. \ref{sec_results} we investigate the dynamics of the systems. In particular we concentrate on how the non-Markovianity affects the recover of coherence in the single-spin system as well as the nonmonotonic entanglement in two-spin system. We summarize in Sec. \ref{sec_summary}.

\section{The quantum jump method}
\label{sec_method}
\subsection{Markovian quantum jump}
\label{sec_mqj}
In Markovian case, the time-evolution of the system's density matrix is governed by the so-called Lindblad master equation (set $\hbar=1$ hereinafter),
\begin{equation}\label{1}
\dot{\rho}(t) = -i[\hat{H}_s, \rho(t)] + \Delta \left( \hat{C} \rho(t) \hat{C}^\dagger - \frac{1}{2} \{ \hat{C}^\dagger \hat{C}, \rho(t) \} \right),
\end{equation}
where $\rho$ is the system's density matrix, $\hat{H}_s$ is the Hamiltonian of the system, $\Delta$ is the constant decay rate of a certain dissipative channel, and $\hat{C}$ is the jump operator representing the dissipative effects on the system. For simplicity, we assume that the jump operator $\hat{C}$ is time-independent throughout the paper. Additionally, for Markovian case the decay rate is positive that guarantees the complete positivity of the corresponding the dynamical map. Eq. (\ref{1}) can be recast as
\begin{equation}
\dot{\rho}(t) = -i\left(\hat{H}_\text{eff}\rho(t)-\rho(t)\hat{H}^\dagger_\text{eff}\right)+\Delta\hat{C}\rho(t)\hat{C}^\dagger,
\label{MarkovianLindblad}
\end{equation}
The first term of the r.h.s. of Eq. (\ref{MarkovianLindblad}) describes the deterministic evolution governed by the effective non-Hermitian Hamiltonian
\begin{equation}\label{effHam}
\hat{H}_{\text{eff}} = \hat{H}_s - i\frac{\Delta}{2} \hat{C}^\dagger\hat{C},
\end{equation}
while the second term describes the stochastic quantum jumps induced by the jump operator $\hat{C}$.

If there is no quantum jump within $\delta t$ and neglecting the higher orders of $\delta t$, the state evolves deterministically as follows,
\begin{equation}
\label{3}
|\psi(t)\rangle\rightarrow|\psi(t + \delta t)\rangle = \left( 1 - i\hat{H}_{\text{eff}}\delta t \right) |\psi(t)\rangle,
\end{equation}
The resulting state must be normalized by $|\psi(t + \delta t)\rangle / \left|  |\psi(t + \delta t)\rangle \right|| $.

If a quantum jump occurs, the state will change discontinuously under the action of the jump operator,
\begin{equation}\label{5}
|\psi(t + \delta t)\rangle \rightarrow \frac{\hat{C}|\psi(t)\rangle}{\left\|\hat{C}|\psi(t)\rangle \right\|}
\end{equation}
with the probability
\begin{equation}\label{6}
p(t) = \Delta \delta t \langle \psi(t) | \hat{C}^\dagger \hat{C} | \psi(t) \rangle.
\end{equation}

The Markovian quantum jump method proceeds by statistically averaging over many independent realizations, yielding the evolution of the system's density matrix as a weighted average over the states of each realization,
\begin{align}\label{8}
\rho(t + \delta t) &= [1 - p(t)]\times
\frac{|\phi(t + \delta t)\rangle \langle \phi(t + \delta t)|}{\langle \phi(t + \delta t) | \phi(t + \delta t)\rangle \ } \nonumber \\
&\quad + p(t)\times
\frac{\hat{C} |\psi(t)\rangle \langle \psi(t)|\hat{C}^\dagger}{\langle \psi(t) | \hat{C}^\dagger \hat{C} | \psi(t) \rangle}.
\end{align}

It should be noted that $\Delta>0$ in Markovian case implies a unidirectional flow of information from the system to the environment.

\subsection{Non-Markovian quantum jump}
When the system couples to a structured environment, the memory effect will result in a feedback of information from the environment. The memory effect is featured by the negative decay rate in a time-local quantum master equation. The negative decay rate can be interpreted as the reverse quantum jump to recover the coherence that lost previously \cite{piiloPRL2008,piiloPRA2009}.

Since the decay rate can be either positive or negative in the non-Markovian case, the dynamics of the system is simulated in two manners: (1) in the region with positive decay rate, one implements the simulation as illustrated in Sec. \ref{sec_mqj}; (2) as the decay rate becomes negative the normal quantum jump is suspended, instead the reversed jumps is activated to counteract the effect of decoherence caused by the normal quantum jump. The reversed quantum jump operators are expressed as follows,
\begin{equation}\label{11}
\hat{D}_{j\to0} = |\psi_0(t)\rangle \langle \psi_1(t)|,
\end{equation}
\begin{equation}\label{12}
\hat{D}_{2\to 1} = |\psi_1(t)\rangle \langle \psi_2(t)|,
\end{equation}
where $|\psi_0(t)\rangle$ denotes the state that no quantum jump takes place since the beginning of the evolution, $|\psi_1(t)\rangle$ denotes the state that underwent one normal quantum jump at $t_1$ when the decay rate is positive and then evolves continuously according to Eq. \eqref{3} until $t$. $|\psi_{2}(t)\rangle$ denotes the state, upon $|\psi_1(t)\rangle$, underwent another normal quantum jump at $t_2>t_1$ and then evolves continuously to $t$.

The reversed quantum jump operator $\hat{D}_{1\to0} $ does not only bring the system back to the moment that a normal jump took place but also evolves the no-jump state to the present time $t$ as if the normal jump never occurs. The reversed jump operator $\hat{D}_{2\to 1}$ works similarly but cancels the effect of the last normal quantum jump. It should be noted that the reversed jump operators defined in Eqs. \eqref{11} and \eqref{12} imply the infinite-time memory effect. Because regardless when the last jump took place, the reversed quantum jump always brings the system to the same target state.

The probability of performing a certain reversed quantum jump is given by
\begin{equation}\label{13}
p_{1\rightarrow 0} = \frac{N_{0}}{N_1} |\Delta(t)|\delta t \langle \psi_0(t)| \hat{C}^\dagger \hat{C} | \psi_0(t)\rangle,
\end{equation}
\begin{equation}\label{14}
p_{2\rightarrow 1} = \frac{N_1}{N_{2}} |\Delta(t)|\delta t \langle \psi_1(t)| \hat{C}^\dagger \hat{C} | \psi_1(t)\rangle,
\end{equation}
where $N_0$, $N_1$ and $N_2$ are the numbers of realizations that undergo zero, one and two normal quantum jumps, respectively \cite{piiloPRL2008,piiloPRA2009}.

\subsection{Quantum trajectory and trajectory class}
\label{sec_QTandTC}

\begin{figure}[hbt]
  \includegraphics[width=1\linewidth]{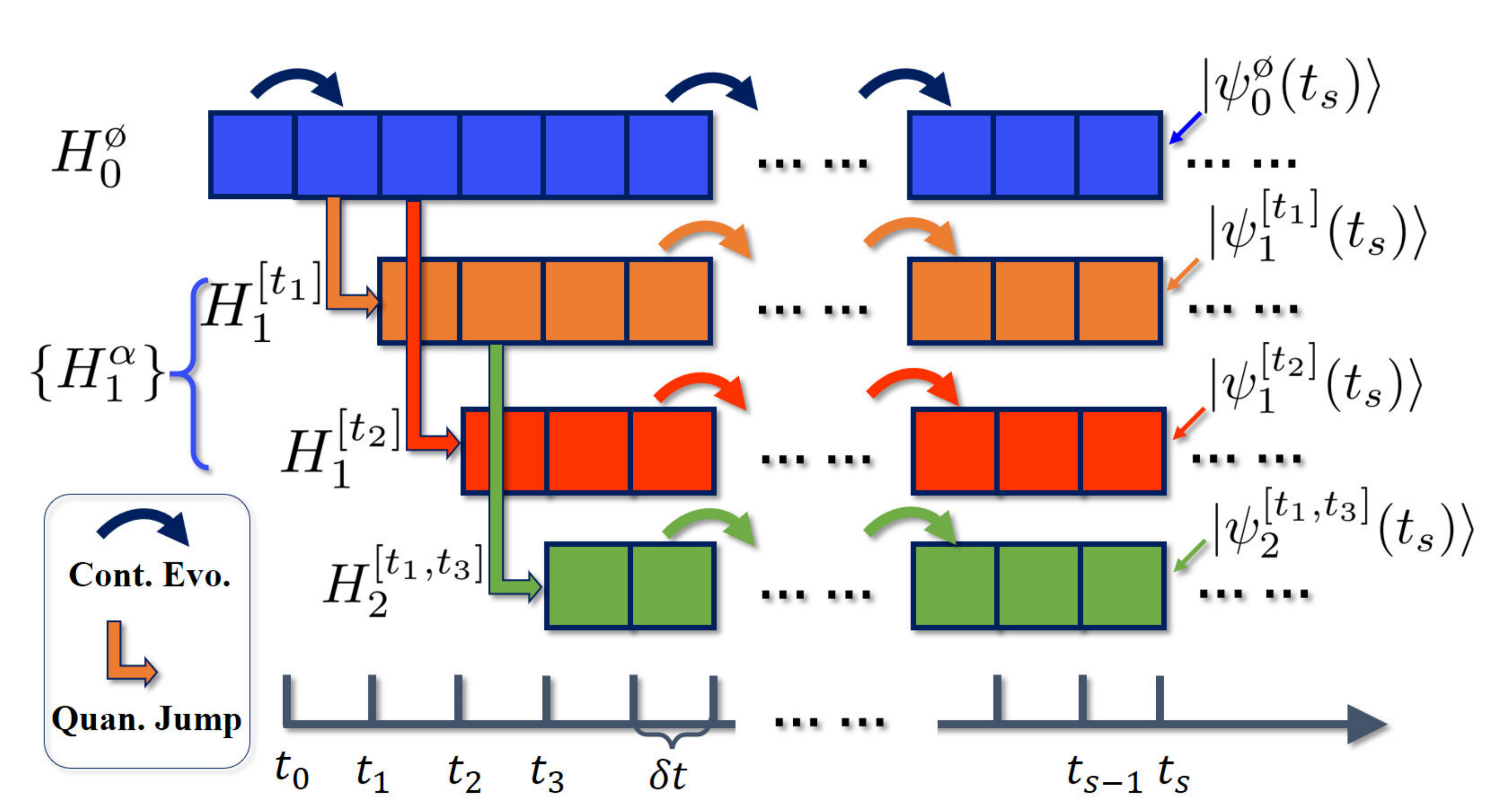}
  \caption{Schematic diagram of the quantum trajectory and trajectory class. The blocks in each row constitute a quantum trajectory $H_n^\alpha$. The neighboring blocks in the same color connected by the curved arrow represents the continuous evolution of the state of the system according to Eq. (\ref{3}). The blocks in different colors connected by the right-angled arrow represents the action of quantum jump after which a new trajectory is created. The blocks in the top row constitute the no-jump quantum trajectory $H_0^{\o}$. The blocks in the second and third rows constitute the quantum trajectories originate from $H_0^{\o}$ and belong to the trajectory class $\{H_1^\alpha\}$. The blocks in the bottom row constitute a trajectory belongs to trajectory class $\{H_2^{[\alpha,t_w]}\}$ with $t_w>t_1$. A reversed quantum jump acting at $t_s$ on the green trajectory will bring the state of the system from $\vert\psi_2^{[t_1,t_3]}(t_s)\rangle$ to $\vert\psi_1^{[t_1]}(t_s)\rangle$.}
  \label{fig_QTandTC}
\end{figure}

In order to implement the NMQJ method in a realistic numerical simulation, the time axis is discretized by short interval $\delta t$ as shown in the Fig. \ref{fig_QTandTC}. The time interval is so short that only one quantum jump could take place within $\delta t$. Suppose that the initial state of the system is $|\psi(t_0)\rangle$. The blocks in the same color constitute a specific evolution path of the system' state, which we refer to as a \textbf{quantum trajectory}, denoted by $H_n^\alpha$. In addition, in a certain quantum trajectory, the quantum state of the system at time $t_s$ is denoted by $|\psi_n^\alpha(t_s)\rangle$. The superscript $\alpha$ is an array of time and represents the time sequence that the normal quantum jump occurred. The subscript $n=\#\alpha$ represents the number of the normal jumps. A quantum trajectory describes the continuous evolution of the system's state solely governed by the effective non-Hermitian Hamiltonian $\hat{H}_{\text{eff}}$ from the latest time point in the time sequence, i.e. $t=\alpha(n)$. Specially, in Fig. \ref{fig_QTandTC} the quantum trajectory in blue describes the evolution of the system's state without any quantum jumps, denoted by $H_n^{\o}$ where $\o$ denotes the empty time sequence. The state of $\hat{H}_n^{\o}$ at time $t_s$ is thus given by
\begin{equation}
\label{psi0_ts}
|\psi_0^{\o}(t_s)\rangle=\frac{\exp{(-i\hat{H}_{\text{eff}}t_s)}|\psi(t_0)\rangle}{||\exp{(-i\hat{H}_{\text{eff}}t_s)}|\psi(t_0)\rangle||},
\end{equation}
where $|\psi(t_0)\rangle$ is the initial state and the denominator is the normalization factor.

We emphasize that according to the definition of quantum trajectory, the initial state of each trajectory is the collapsed state after the jump operator acting on a continuously evolved state. For example,
$H_1^{[t_2]}$ represents the quantum trajectory produced by the action of jump operator in $(t_1,t_2]$ on the trajectory $H_0^{\o}$. The initial state of $H_1^{[t_2]}$ is thus $|\psi_1^{[t_2]}(t_2)\rangle\sim \hat{C}\exp{(-i\hat{H}_{\text{eff}}t_1)}|\psi(t_0)\rangle$ and the final state at $t_f$ is
$|\psi_1^{[t_2]}(t_f)\rangle\sim\exp{[-i\hat{H}_{\text{eff}}(t_f-t_2)]}|\psi_1^{[t_2]}(t_2)\rangle$.

\jj{ }

Next, we introduce the concept of a \textbf{trajectory class}, denoted by $\{H_n^\alpha\}$. The trajectory class is a set of quantum trajectories labeled by the time sequence $\alpha=[\alpha',t_w]$ with the same $\alpha'$ but different $t_w$. The quantum trajectories belonging to $\{H_n^\alpha\}$ originate from the same mother trajectory $H_{n-1}^{\alpha'}$ but are produced at different time $t_w$ by the action of normal quantum jump. For example, the orange and red trajectories in Fig. \ref{fig_QTandTC} belong to the same trajectory class $\{H_1^\alpha\}$ because both of them are generated from the mother trajectory $H_0^{\o}$ in blue via the action of the jump operator $\hat{C}$ at $t_1$ and $t_2$, respectively. In a similar fashion, the orange trajectory can also generate a new trajectory class, denoted by $\{H_2^{[t_1,t_w]}\}$, consists of the sub-trajectories produced by the action of each single quantum jump at any time $t_w>t_1$. Obviously the trajectory class $\{H_0^{\o}\}$ only includes the no-jump trajectory $H_0^{\o}$.

In general, each quantum trajectory can generate a trajectory class by the single-jump acting on it at any time, and conversely, the quantum trajectory itself may also belong to a trajectory class consisting of those trajectories sharing the same mother trajectory but jumped at different time. We would emphasize that the quantum jump takes place instantaneously and we cannot resolve the exact time during the time interval $\delta t$. However, as $\delta t$ being sufficiently short, we can approximate the quantum jump to take place at $t_w$ in the period $(t_{w-1},t_w]$. Moreover, $\delta t$ is so short that only one jump can take place in the time interval.

The motivation of introducing the quantum trajectory class is to group the trajectories with the same `memory' together. By the same memory, we mean that action of a reversed quantum jump will bring these trajectories back to their common mother trajectory, provided the memory time is infinite long, i.e. the reversed quantum jump transforms $\{H_{n}^{[\alpha',t_w]}\} \rightarrow H_{n-1}^{\alpha'}$. As will be seen soon, this classification will facilitate in calculating the probability of the occurrence of a reversed quantum jump in the region of negative decay rate.

\subsection{Existence probability of quantum trajectory}
\label{sec_EP}
In a conventional numerical simulation with quantum trajectories, the state of the system is sampled by a large number of realizations. In each realization, starting from the same initial state, the system evolves stochastically and the state of the quantum system at arbitrary time $t_w$  is obtained by averaging the states over all the realizations.

Suppose that the total number of realizations is $N$ and the number of realizations that stay in the trajectory $H_n^\alpha$ at time $t_w$ is $N_n^\alpha(t_w)$. In the limit of infinite $N$, the existence probability of each trajectory is defined as
\begin{equation}
\label{ep_Kn}
K_n^\alpha(t_w)=\lim_{N\rightarrow\infty}{\frac{N_n^\alpha(t_w)}{N}}.
\end{equation}
The existence probability in Eq. (\ref{ep_Kn}) represents the proportion of the state in a certain trajectory among all the possible states. It can be used to construct the density matrix of the system as follows,
\begin{equation}
\rho(t) = \sum_{n}\sum_{\alpha}{K_n^{\alpha}\lvert\psi_n^{\alpha}(t)\rangle\langle\psi_n^\alpha\rvert}.
\label{dm_construction}
\end{equation}
The unity trace of the density matrix is satisfied by definition.

Now we are in the position to calculate the existence probability of a certain quantum trajectory. We will discuss in the cases of positive and negative decay rates respectively.
\subsubsection{$\Delta(t)>0$ case}
Let us first focus on the time duration $[t_0,t_P]$ in which the decay rate is positive $\Delta(t)>0$.
To start, we employ the result reported in Ref. \cite{piiloPRL2008}, the existence probability of quantum trajectory (class) $H_0^{\o}$ at $t_w\in[t_0,t_P]$ is given by
\begin{equation}
\label{eq_K00}
K_0^{\o}(t_w) = \prod_{s=1}^{w-1} \left[1 - p^{\o}_0(t_s)\right],
\end{equation}
where
\begin{equation}
\label{eq_p00}
p^{\o}_0(t_s) = \Delta(t_s) \delta t \langle \psi_0(t_s) \lvert \hat{C}^\dagger \hat{C} \rvert \psi_0(t_s) \rangle,
\end{equation}
represents the probability of the normal quantum jump occurs at time $t_s$. The continuously evolved (without any quantum jump) state $|\psi_0(t_s)\rangle$ is given in Eq. (\ref{psi0_ts}).

Based on Eq. (\ref{eq_K00}), the existence probability of the quantum trajectories undergoes only one jump at arbitrary $t_w$, i.e. $H_1^{\alpha}$ with $\alpha=[t_w]$, can be computed as the following. For a single two-level system, if the state after a quantum jump is the eigenstate of the effective Hamiltonian, the system will stay in the lower level and no more quantum jump can take place within the new trajectory. Thus the existence probability yields
\begin{equation}\label{20}
K_{1}^{[t_w]}(t_P) =  K_0^{\o}(t_{w-1}) p^{\o}_0(t_w),
\end{equation}
which is the probability of a quantum jump occurring at $t_w$ conditioned on that no jump takes place before $t_w$ \cite{piiloPRA2009}. In Eq. (\ref{20}) the time sequence has been expressed by the single-element array $[t_w]$ for clarity.

However if extended to the many-body system or the driven single-qubit system, the quantum state after a (local) quantum jump is usually not an eigenstate of the non-Hermitian Hamiltonian (\ref{effHam}). As a consequence the collapsed state will be excited by the effective driving in the many-body system or the external driving in the single-body system, modifying Eq. (\ref{20}) as follows,
\begin{equation}\label{305}
K_{1}^{[t_w]}(t_{P})=K_{0}^{\o}(t_{w-1})p^{\o}_{0}(t_{w}) \prod_{s=w+1}^{M-1}\left[1-p^{[t_w]}_{1}(t_{s})\right],
\end{equation}
where
\begin{equation}\label{19}
p^{[t_w]}_{1}(t_{s}) = \Delta(t_s)\delta t\langle \psi^{[t_w]}_1(t_s) \lvert \hat{C}^\dagger \hat{C} \rvert \psi^{[t_w]}_1(t_s) \rangle
\end{equation}
is the probability of the quantum jump taking place at $t_s > t_w$ with respect to the state $|\psi^{[t_w]}_1(t_s)\rangle$. Here  $|\psi^{[t_w]}_1(t_s)\rangle$ is the state after the normal quantum jump at $t_w$ and continuously evolved to $t_s$ governed by the effective Hamiltonian.

As a consequence, for a generic quantum trajectory which undergoes $n$ quantum jumps along time sequence $\alpha$, the existence probability can be obtained iteratively as
\begin{equation}
K_n^\alpha(t_P) = K_{n-1}^{\alpha'}(t_{w-1})p_{n-1}^{\alpha'}(t_w)\prod_{s=w+1}^{P-1}\left[1-p^{\alpha}_{n}(t_{s})\right],
\end{equation}
with $\alpha=[\alpha',t_w]$. $K_n^\alpha(t_P)$ is the product of the existence probability of a quantum trajectory underwent $(n-1)$ quantum jumps before $t_{w-1}$, the probability of a quantum jump takes place at $t_w$,  and the probability of no more jump takes place in the future $t>t_w$. The probability of a quantum jump occurs at $t$ with respect to the state $|\psi_n^\alpha(t)\rangle$ is
\begin{equation}
p_n^\alpha(t) = \Delta(t)\delta t\langle \psi^{\alpha}_n(t) \lvert \hat{C}^\dagger \hat{C} \rvert \psi^{\alpha}_n(t) \rangle.
\label{prob_pn}
\end{equation}

\subsubsection{$\Delta(t)<0$ case}
When the decay rate becomes negative, the reversed quantum jump is switched on and it may bring a quantum trajectory back to its mother trajectory. On the other hand since the normal quantum jump is suspended no new quantum trajectory will be created.

In the $\Delta(t)<0$ region, the existence probability of trajectory $H_0^{\o}$ increases because the trajectories belong to the trajectory class $\{H_1^\alpha\}$ may be transferred to $H_0^{\o}$ through a reversed quantum jump. For instance, in the first time interval of the $\Delta<0$ region, the existence probability of trajectory $H_0^{\o}$ is modified as follows,
\begin{equation}\label{222}
K_0^{\o}(t_{P} + \delta t) = K_0^{\o}(t_P) + \sum_{\alpha}{K_{1}^{\alpha}(t_P)q_{0}^{\o}(t_P)}.
\end{equation}
The first term of the r.h.s. of Eq. (\ref{222}) represents the existing probability at the end of the $\Delta>0$ region, while the second term represents the contribution of trajectory class $\{H_1^\alpha\}$ through the reversed quantum jump. The probability of a reversed jump occurring at $t_P+\delta t$ is given by
\begin{equation}\label{233}
q_{0}^{\o}(t_P)=\frac{N_0^{\o}(t_P)}{\sum_\alpha{ N_1^{\alpha}(t_P)}} \lvert \Delta(t_P) \rvert \delta t \langle \psi_0^{\o}(t_P) \lvert \hat{C}^\dagger \hat{C} \lvert \psi_0^{\o}(t_P) \rangle.
\end{equation}

One can see that the probability of a reversed quantum jump taking place in each quantum trajectory belonging to $\{H_1^\alpha\}$ depends on (i) the population with respect to the target state $|\psi_0^{\o}(t_P)\rangle$; (ii) the ratio of the realization numbers of no-jump trajectory and all the single-jumped trajectory, this is due to the effect of infinite memory time. Therefore the proportion of each trajectory multiplies the corresponding $q^{\o}_0(t_P)$ followed by the sum over all the possible trajectory yields all the contributions to the increment of $K_0^{\o}(t_{M} + \delta t)$.

Recall the definition of existence probability, one has
\begin{equation}\label{244}
\frac{N_0^{\o}(t_P)}{\sum_\alpha N_1^{\alpha}(t_P)} =  \frac{N_0^{\o}(t_P)/N}{\sum_\alpha N_1^{\alpha}(t_P)/N}= \frac{K_0^{\o}(t_P)}{\sum_{\alpha} K_{1}^{\alpha}(t_P)}.
\end{equation}
Substitute Eqs. \eqref{eq_K00}, \eqref{233} and \eqref{244} into Eq. \eqref{222}, and recast the probability of the reverse quantum jump in Eq. (\ref{233}) as $q_{0}^{\o}(t_P)=-\frac{N_0^{\o}(t_P)}{\sum_\alpha{ N_1^{\alpha}(t_P)}}p_0^{\o}(t_P)$, the existence probability of $H_0^{\phi}$ yields
\begin{eqnarray}\label{25}
K_0^{\o}(t_P + \delta t)& = &K_0^{\o}(t_P) + K_0^{\o}(t_P) q_0^{\o}(t_P)\cr\cr
&=&K_0^{\o}(t_P)\left[1 - p_0^{\o}(t_P)\right]\cr\cr
&=&\prod_{s=0}^{M} \left[1 - p_0^{\o}(t_s)\right],
\end{eqnarray}
where $p_0^{\o}(t_s)$ is given by Eq. (\ref{eq_p00}).

Iterating Eq. (\ref{25}) through out the whole $\Delta(t)<0$ region $(t_P,t_{M'}]$, the existence probability of no-jump trajectory at any moment $t_w$ is given as follows,
\begin{equation}\label{277}
\begin{aligned}
K_0^{\o}(t_{w}) &= \prod_{s=0}^{w-1} \left[1 - \Delta(t_w)\delta t  \langle \psi_0^{\o}(t_s) \lvert \hat{C}^\dagger \hat{C} \lvert \psi_0^{\o}(t_s) \rangle\right].
\end{aligned}
\end{equation}

One can see that the existence probability $K_0^{\o}(t_s)$ at any moment can be calculated straightforwardly with the help of the continuously evolved state $|\psi_0^{\o}(t_s) \rangle\sim \exp{(-i\hat{H}_{\text{eff}}t_s)|\psi(t_0)\rangle}$ regardless of the sign of the decay rate. Notice that the existence probability $K_0^{\o}(t_s)$ should not exceed unity by definition. When $K_0^{\o}(t_s)$ reaches to unity, all the other trajectories are brought back to the identical trajectory $H_0^{\o}$ and are frozen in this trajectory until the decay rate becomes positive, the normal quantum jump is switched on again.

Next, we derive the  expression for the existence probability of the single-jump quantum trajectory $H_1^\alpha$. Unlike $H_0^{\o}$, the existence probability $K_1^{\alpha}(t_P+\delta t)$ may decrease due to the probability transferring away to $K_0^{\o}(t_P+\delta t)$ or increase due to the acceptance of probability from $K_2^{\alpha'}(t_P+\delta t)$ with $\alpha'=[\alpha,t_w]$. Notably, here it is assumed that reduction amount of each $K_1^\alpha(t_P)$  has equal probability to be transferred away to $K_0^{\o}$ because the memory time is infinite throughout the entire time evolution. Therefore the expression for $K_1^\alpha(t_P+\delta t)$ is given as follows,
\begin{equation}
\label{K1tM1}
K_1^\alpha(t_P+\delta t)=K_1^\alpha(t_P)\left[1-q^{\o}_0(t_P)\right]+\sum_{\alpha'}{\left[K_2^{\alpha'}(t_P)q_1^\alpha(t_P)\right]},
\end{equation}
where
\begin{eqnarray}
\label{2331}
q_1^\alpha(t_P)&=&\frac{N_1^\alpha(t_P)}{\sum_{\alpha'}{N_2^{\alpha'}(t_P)}}|\Delta(t_P)|\delta t\langle\psi_1^\alpha(t_P)|\hat{C}^\dagger\hat{C}|\psi_1^\alpha(t_P)\rangle\cr\cr
&=&-\frac{N_1^\alpha(t_P)}{\sum_{\alpha'}{N_2^{\alpha'}(t_P)}}p_1^{\alpha}(t_P),
\end{eqnarray}
stands for the probability that a two-jump trajectory $H_2^{\alpha'}$ may jump back to $H_1^\alpha$ at $t_P$, which depends on the population with respect to the state of the target trajectory $|\psi_1^{\alpha}(t_P)\rangle$. Eq. (\ref{K1tM1}) possesses a quite intuitive physical meaning: The first term of the r.h.s. of Eq. (\ref{K1tM1}) represents the reduction of $K_1^\alpha(t_P)$ due to the reversed jump to its mother trajectory $H_0^{\o}$, while the second term represents the increment of $K_1^\alpha(t_P)$ due to the contributions via the reversed jump of all the trajectories belonging to class $\{H_2^{\alpha'}\}$ . Substituting Eqs. (\ref{233}) and (\ref{2331}) into Eq. (\ref{K1tM1}), one can obtain
\begin{eqnarray}
K_1^\alpha(t_P+\delta t)&=&K_1^\alpha(t_P)\left[1-p_1^\alpha(t_P)\right]\cr\cr
&&+\frac{K_0^{\o}(t_P)K_1^\alpha(t_P)}{\sum_\alpha{K_1^\alpha(t_P)}}p_0^{\o}(t_P).
\end{eqnarray}

Following the procedure in deriving Eq. (\ref{K1tM1}), one can obtain the updated existence probability $K_n^\alpha(t_P+\delta t)$ for a generic quantum trajectory as follows,
\begin{eqnarray}
K_n^\alpha(t_P+\delta t)&=&K_n^\alpha(t_P)\left[1-p_{n}^{\alpha}(t_P)\right]\cr\cr
&&+\frac{K_{n-1}^{\alpha''}(t_P)K_n^\alpha(t_P)}{\sum_{\alpha}{K_n^\alpha(t_P)}}p_{n-1}^{\alpha''}(t_P),
\label{generic_traj}
\end{eqnarray}
with $\alpha = [\alpha'',t_w]$. The $K_{n-1}^{\alpha''}$ denotes existence probability of the mother trajectory of $H_n^{\alpha}$.

It is interesting that although in the $\Delta(t)<0$ region the existence probability of a generic trajectory may increase due to the reversed quantum jump from its sub-trajectories, the detailed information of these sub-trajectories is not required. As shown in Eq. (\ref{generic_traj}), the updated $K_n^{\alpha}$ only depends on the existence probabilities of its mother trajectory and itself, as well as the probabilities of quantum jump (\ref{prob_pn}) with respect to the states of its mother trajectory and itself. This enables us to implement the calculation in a top-bottom manner, i.e., one first calculates $K_0^{\o}$, then the $K_1^{\alpha}$s, and so on and so forth.

Moreover the calculation can be truncated at a certain hierarchy $n^*$ if the existence probability of trajectory class $\{H_n^\alpha\}$ with $n>n^*$ is below a threshold. Indeed, as will be seen soon, in the presented models in Sec. \ref{sec_models}, the quantum trajectory $H_0^{\o}$ and the trajectory class $\{H_1^{\alpha}\}$ occupy almost all the portion of the possible quantum trajectories in a realistic example. Therefore, in such situations, we may safely neglect the contributions of the trajectories those undergo more than two quantum jumps. The computational cost in storing the history of time-evolution is dramatically reduced. If the existence probability of the quantum trajectory that experiences multiple quantum jumps is non-negligible, the contribution must be taken into account in the computation and therefore a higher truncation number $n^*$ is needed.

Additionally, the normalization of the existence probability $\sum_{n,\alpha}K_n^\alpha(t)=1$ is satisfied by definition if all the possible trajectories are taken into account. However, in a realistic implementation with truncation order $n^*$, the increment of $K_{n^*}^{\alpha}$ should be excluded.

\section{The models}
\label{sec_models}
\begin{figure}[hbt]
  \includegraphics[width=0.95\linewidth]{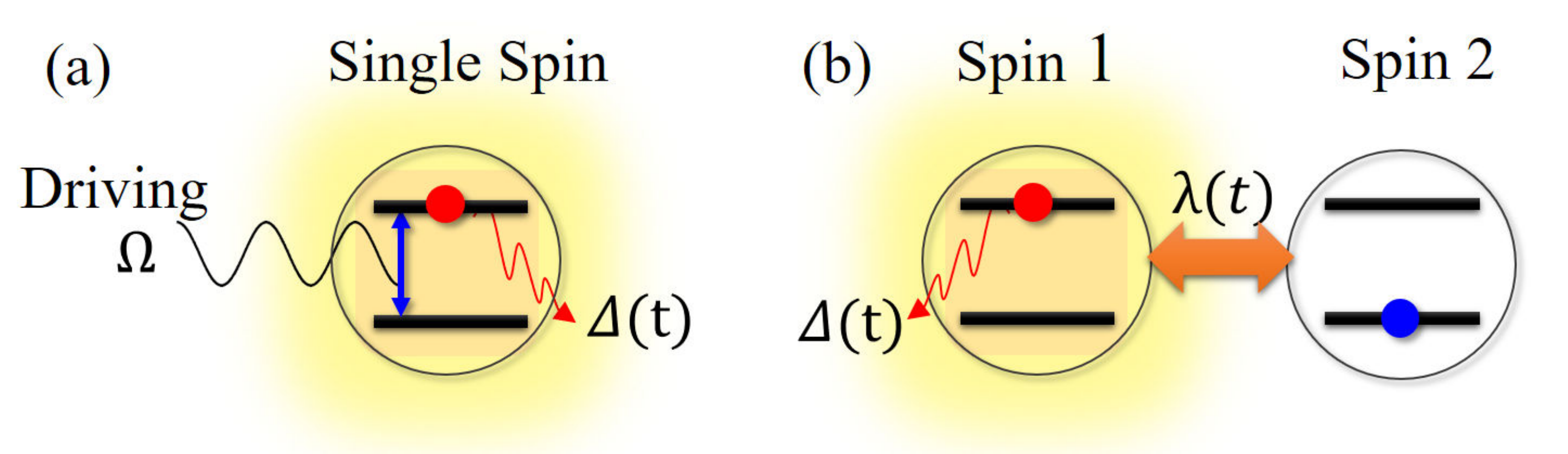}
  \caption{The models. (a) Model I: a spin-1/2 system with an external driving $\Omega$ is subjected to an structured reservoir, leading to a decay rate $\Delta(t)$. (b) Model II: two spins interact with each other with the coupling strength $\lambda(t)$. One of the qubits is subjected to the reservoir while the other one is isolated from the environment. The spectral densities of the reservoirs in both models are Lorentzian.}
  \label{models}
\end{figure}

In order to check the performance, we are going to apply the modified NMQJ method to two toy models. We restrict the discussion on the dynamics of spin-1/2 systems (with two levels $\vert\uparrow\rangle$ and $\vert\downarrow\rangle$) subject to noisy environment as shown in Fig. \ref{models}. Here we assume that the temperature of the environment is zero such that the thermal excitation can be neglected \cite{luoma_PRL2020}. Thus the environment can only induce the deexcitation of the spin through the incoherent flip down to the $z$-direction.

In Model I we consider a driven single spin-1/2 system with incoherent spin flip induced by the environment, yielding the time-local master equation as
\begin{equation}
\label{model_i}
\dot{\rho}=-i[\hat{H}_{\text{I}},\rho] + \Delta(t)(\hat{\sigma}^-\rho\hat{\sigma}^+ -\frac{1}{2} \{\hat{\sigma}^+\hat{\sigma}^-,\rho\}),
\end{equation}
where $\hat{H}_{\text{I}}=\Omega\hat{\sigma}^x$ presents an external driving field imposing on the spin along $x$-direction, $\Omega$ is the Rabi frequency, $\hat{\sigma}^\alpha$ with $\alpha=x,y,z$ are Pauli matrices for spin-1/2 system and the raising and lowering operators are $\hat{\sigma}^{\pm}\equiv (\hat{\sigma}^x\pm i\hat{\sigma}^y)/2$. The notation $\{\cdot,\cdot\}$ stands for the anti-commutator.

In Model II, we consider a coupled two-spin system with spin interaction along $x$-direction. One spin (labeled by 1) is subjected to an environment leading to incoherent spin-flip and the other spin (labeled by 2) is isolated from the environment. The dynamics of such two-spin system is described by the time-local master equation as follows,
\begin{equation}
\label{model_ii}
\dot{\rho}=-i[\hat{H}_{\text{II}},\rho ]+ \Delta(t)(\hat{\sigma}_1^-\rho\hat{\sigma}_1^+ - \frac{1}{2}\{\hat{\sigma}_1^+\hat{\sigma}_1^-,\rho\}),
\end{equation}
where $\hat{H}_{\text{II}}=\lambda(t)\hat{\sigma}^x_1\hat{\sigma}^x_2 $, and $\lambda(t)$ is the coupling strength between the spins.

The jump operators for both models are the lowering operator that flips the spin down to the $z$-direction. However the spin-down state is neither the eigenstate of $\hat{H}_1$ nor $\hat{H}_2$, thus the state after a quantum jump may still be excited through the continuous evolution governed by the effective non-Hermitian Hamiltonian.

Here we assume the dissipative spins (the single spin in Model I and the spin 1 in Model II) interact with a structured environment. The spectral density of the environment is a Lorentzian with center frequency around $\omega$ and width $\Gamma$,
\begin{equation}\label{47}
J_{\text{Lor}}(\nu) = \frac{\eta^{2}}{2\pi}\frac{\Gamma^2}{(\nu-\omega)^{2}+\Gamma^{2}},
\end{equation}
where $\eta$ denotes the coupling strength between the system and the environment. The spectral density of the quantized electromagnetic field inside an imperfect cavity is well approximated by the the spectrum in Eq. (\ref{47}). The width of the distribution quantifying the leakage of photons through the cavity mirrors is characterized by the parameter $\Gamma$.
When characteristic time scale $\tau_S$ is much shorter than the reservoir correlation time $\tau_C$, one can make the secular approximation on the system-environment interaction, leading to the following time-dependent decay rate in the time-local Lindblad master equations (\ref{model_i}) and (\ref{model_ii}),
\begin{equation}\label{44}
\Delta(t) = \frac{\eta^2\{1+e^{-\Gamma t}\left[q_0\sin{(q_0\Gamma t)-\cos{(q_0\Gamma t)}}\right]\}}{2(1+q_0^2)},
\end{equation}
where $q_0$ denotes the detuning of the central frequency of the Lorentzian spectrum to the frequency of two-level system \cite{haikka_PhysScr_2010}. The limit of weak coupling is reached when $\eta^2$ is smaller than the smallest relevant frequency in the system. Here in the rest of the paper we work in units of $\Gamma$ and set $\eta=10$ and $q_0=6$ in the numerical computation. The time intervals in both models are chosen as $\delta t=10^{-3}\Gamma^{-1}$.

\begin{figure}[htpb]
	\centering
	\includegraphics[width=0.85\linewidth]{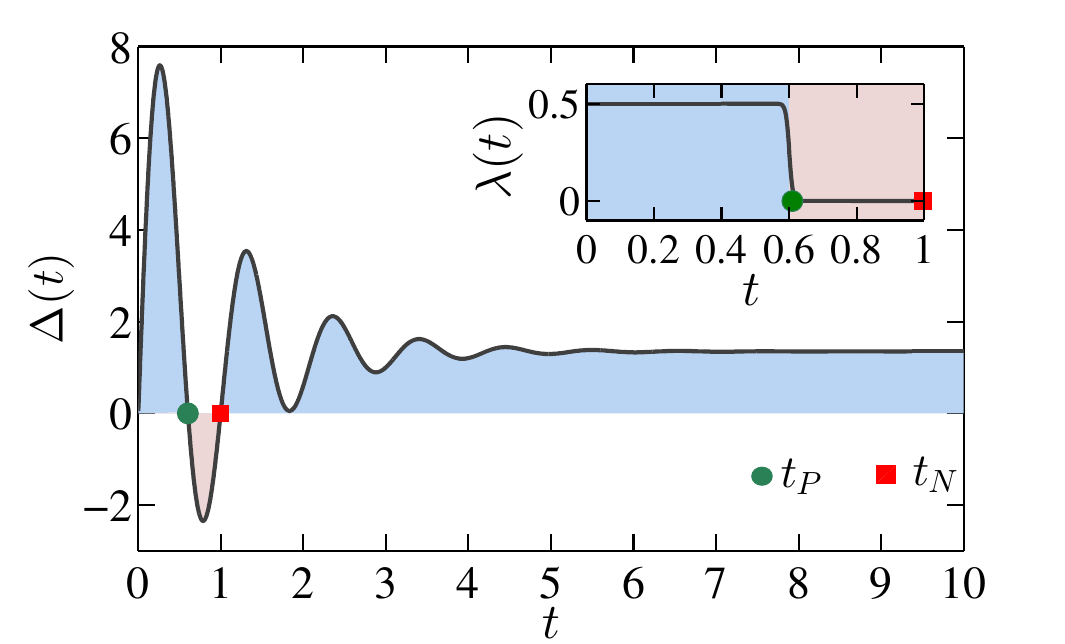}
	\caption{Main panel: the time-dependent decay rate $\Delta(t)$ induced by the reservoir with Lorentzian spectral density in both models. The area shaded in blue marks the $\Delta(t)>0$ region and the area shaded in pink marks the $\Delta(t)<0$ region. The parameters are chosen as $\eta=10$ and $q_0=6$ in Eq. (\ref{44}). The time is in units of $\Gamma^{-1}$. The inset: the time-dependent spin coupling $\lambda(t)$ in case (ii) of Model II. For $t<t_P$, $\lambda(t)\approx0.5$; for $t_p<t<t_N$, $\lambda(t)\approx0$.}
	\label{decayrate}
\end{figure}

In Fig. \ref{decayrate}, it is shown the decay rate of the spin for both models. The decay rate is positive in the initial stage $t<t_P$ and becomes negative in a intermediate stage $t_P<t<t_N$, then becomes positive for $t>t_N$ and approaches the stationary value in the long-time limit. The $t_P$ and $t_N$ denote the end of the first positive and the negative decay regions, respectively. The non-Markovian dynamics only appears in the early time of the evolution, so we will mainly concentrate on the dynamics by the end of the negative decay rate region, $t\le t_N$.

\section{Results}
\label{sec_results}
\subsection{Model I: revival of coherence}
Let us first investigate the dynamics of the single spin-1/2 system in Model I. The Rabi frequency of the driving field is chosen as $\Omega = 0.5$. The spin precesses around the external field along $x$-direction which leads to a coherent oscillation between the $\vert\uparrow\rangle$ and $\vert\downarrow\rangle$ states. The normal quantum jump always incoherently flips the spin to the state $\vert\downarrow\rangle$. The jump probability is proportional to the population of the $\vert\uparrow\rangle$ state. Therefore the existence probability of the quantum trajectories depends on the initial state of the spin. We characterize the initial state via the polar and azimuthal angles $(\theta,\phi)$ in the Bloch sphere.

\begin{figure}[htpb]
	\centering
	\includegraphics[width=0.9\linewidth]{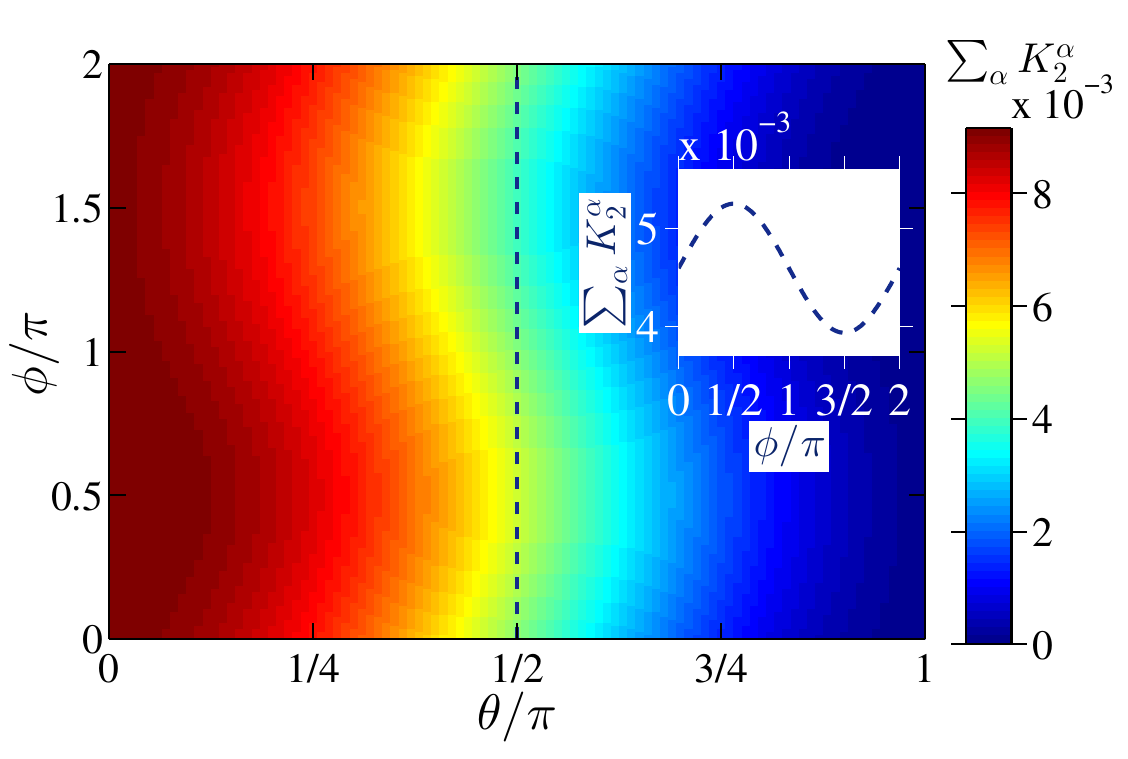}
	\caption{The total existence probability of quantum trajectories undergo two quantum jumps at the end of the positive decay rate region $t=t_P$ in the $\theta-\phi$ plane. The inset shows the dependence of $\sum_{\alpha}{K_2^\alpha(t_P)}$ on the azimuthal angle $\phi$ for $\theta=\frac{\pi}{2}$, corresponding to the vertical dashed line in the main panel.}
	\label{K2_ini_1}
\end{figure}

In order to determine the truncated number of quantum jumps, we check the the total existence of probability of quantum trajectories undergoes two quantum jumps at the end of the positive decay region. The value of $\sum_{\alpha}{K_2^\alpha(t_P)}$ with various initial states are shown in Fig. \ref{K2_ini_1}.  For the initial states on the equator, the maximal $K_2(t_P)$ locates at $\phi=\frac{\pi}{2}$ corresponding to the spin polarizing along the positive $y$-direction. This is because the precession brings the Bloch vector passing by the north pole leading to more probability of normal quantum jump. Oppositely the initial state along negative $y$-direction will result in less $K_2(t_P)$ because the precession is counterclockwise, both the coherent and incoherent spin-flipping tend to reduce the population of $\vert\uparrow\rangle$ state thus suppress the probability of normal quantum jump. It is also worth noting that the magnitude of $K_2(t_P)$ in the whole $\theta-\phi$ plane is less than $10^{-2}$, indicating the contribution of the two-jump trajectories is so small that can be neglected in the computation.

\begin{figure*}[htpb]
	\centering
	\includegraphics[width=0.8\linewidth]{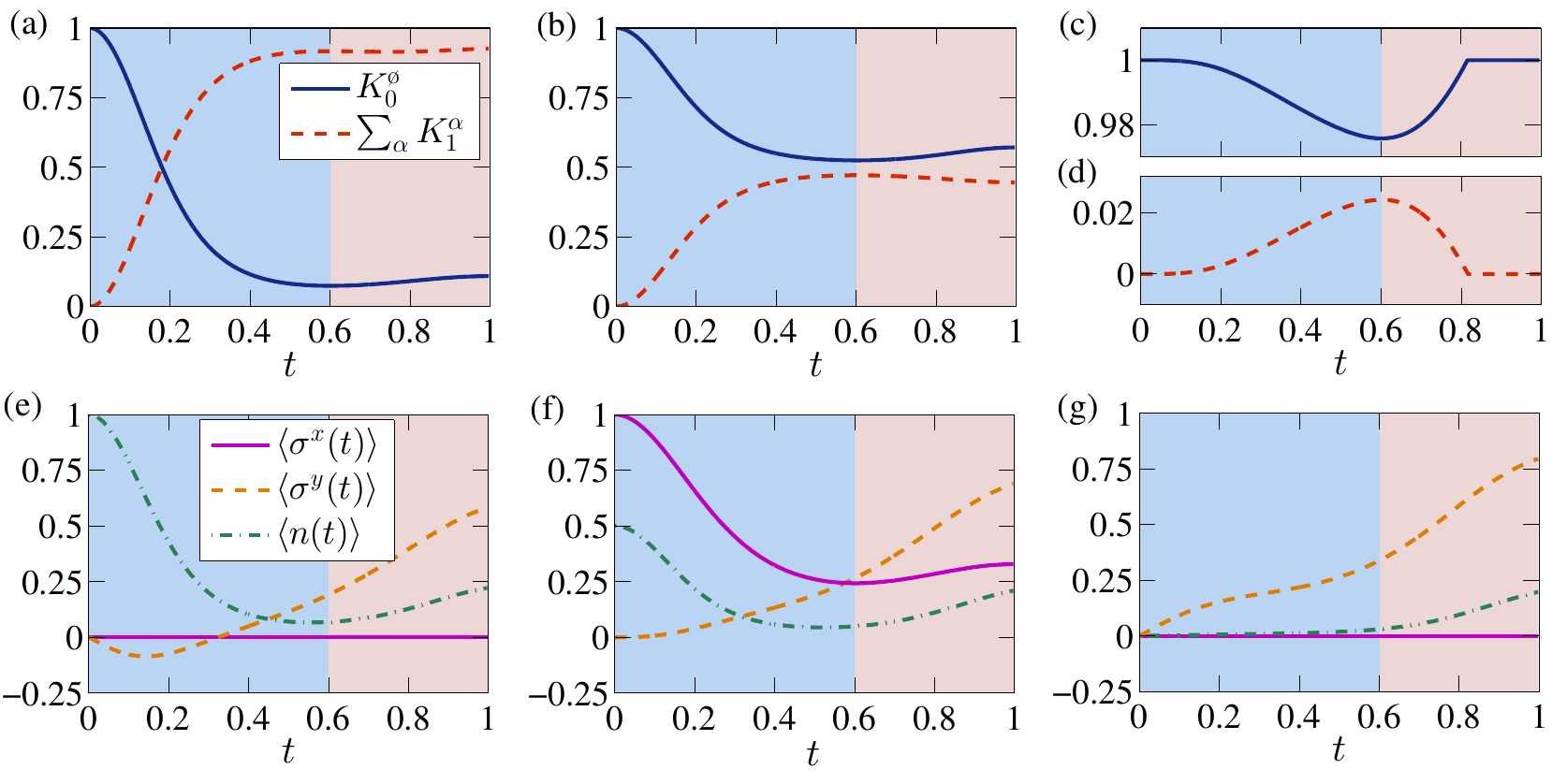}
	\caption{The dynamics of the driven single spin in Model I. The top panels show the time-evolution of the $K_0^{\o}(t)$ and $\sum_\alpha{K_1^\alpha}(t)$ with initial states $|\uparrow\rangle$ (a), $\frac{1}{\sqrt{2}}(|\uparrow\rangle+|\downarrow\rangle)$ (b) and $|\downarrow\rangle$ (c). The bottom panels show the time-evolution of the components of Bloch vectors. The initial states in (e) - (f) correspond to those in (a) - (c). The driving Rabi frequency is chosen as $\Omega = 0.5$ and the decay rate is as given in Fig. \ref{decayrate}.}
	\label{TLS1}
\end{figure*}

In Fig. \ref{TLS1} more details of the time-evolution of the single spin for $t\le t_N$ are shown with various initial states: the spin-up state, spin-down state and the superposition of both. As mentioned above, since the $K_2$ is very small we only consider the trajectories $H_0^{\o}$ and $\{H_1^{\alpha}\}$. The upper panels of Fig. \ref{TLS1} shows the time-dependence of $K_0^{\o}(t)$ and the total existence probability of all the trajectories with one-jump $K_1(t)=\sum_\alpha{K_1^\alpha(t)}$. In the Markovian region (blue shaded), because the normal jumps occur
the $K_0^{\o}$ decreases monotonically while $K_1(t)$ increases. Moreover, from Figs. \ref{TLS1}(a)-(c) one can see that as the initial population of the $\vert\uparrow\rangle$ decreasing, the system is less likely to jump. The lower panels show the time-evolution of the components of the Bloch vector. Because the external driving is along $x$-direction, if the initial state lies in the $y$-$z$ plane the $\hat{\sigma}^x$ component is always zeros, as shown in Figs. \ref{TLS1}(e) and (g).

When the decay rate becomes negative (the pink shaded area in Fig. \ref{TLS1}), the reversed quantum jump is activated and the coherence of the system is recovered partly. This is manifested by the revival of $K_0^{\o}$ in the non-Markovian region. In particular, for the case of spin-down initial state, the $K_0^{\o}$ reaches again the unity before the end of the negative decay rate region, implying that the all trajectories are brought back to the $H_0^{\o}$ trajectory by the reversed quantum jump. The dynamics can be described by the non-unitary evolution governed by the effective non-Hermitian Hamiltonian (\ref{effHam}).

\subsection{Model II: Sudden death and revival of entanglement}
Now let us investigate the dynamics of the two spins in model II. The Hamiltonian is the spin interaction along $x$-direction and generates a global operation acting on both the spins, while the dissipation acts on the spin 1 locally. We consider two cases of the coupling strength $\lambda(t)$: (i) the constant $\lambda(t) = \lambda_0=0.5$; (ii) the time-dependent $\lambda(t)=\lambda_0-\lambda_0\{1+\exp{[-2\beta (t-t_P)]}\}^{-1}$ with $\beta=100$. The coupling strength in case (ii) equals to $\lambda_0$ for $\Delta(t)>0$ and vanishes rapidly as $\Delta(t)$ becomes negative.

\begin{figure}[htpb]
	\centering
	\includegraphics[width=1\linewidth]{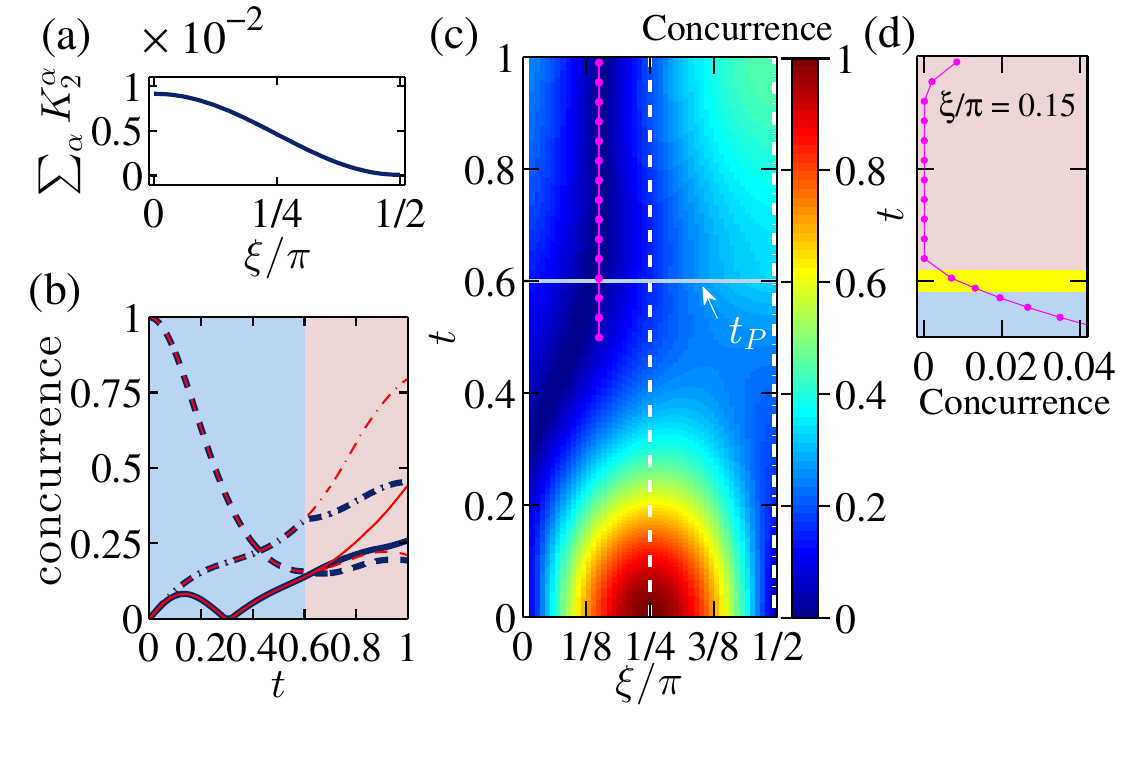}
	\caption{(a) The total existence probability of quantum trajectories with two jumps at the end of positive decay rate region $t=t_P$ for various initial state parameterized by $\xi$. (b) The thin and thick lines show the time-evolution of concurrence with the coupling strength $\lambda(t)$ in case (i) and case (ii), respectively. The initial states are $|\uparrow_1\downarrow_2\rangle$ (solid lines), $|\downarrow_1\uparrow_2\rangle$ (dotted-dashed lines) and $\frac{1}{\sqrt{2}}(\uparrow_1\downarrow_2\rangle+|\downarrow_1\uparrow_2\rangle)$ (dashed lines). (c) The time-evolution of concurrence for various initial states. The vertical lines in white correspond to the three specific cases in panel (b), the magenta linked circles corresponds to the case of $\xi/\pi=0.15$ in panel (d), and the horizontal solid line marks the boundary between the positive and negative decay rate regions. (d) The sudden death and revival of entanglement during the time-evolution for the initial state with $\xi/\pi = 0.15$. The area shaded in yellow highlights the tail of nonzero $\lambda(t)$ at the beginning of $\Delta(t)<0$ region.}
	\label{spin2}
\end{figure}

There are two subspaces of the joint Hilbert space $\mathcal {H}_{\text{odd}}=\{\vert\uparrow_1\downarrow_2\rangle,\vert\downarrow_1\uparrow_2\rangle\}$ and $\mathcal {H}_{\text{even}}=\{\vert\uparrow_1\uparrow_2\rangle,\vert\downarrow_1\downarrow_2\rangle\}$. Initializing the system in a state belonging to a certain subspace, the interaction $\hat{H}_{\text{II}}$ will manipulate the state inside the given subspace while the jump operator $\hat{\sigma}^-_1$ will kick the state between the two subspaces. Without loss of generality, in the following discussion we choose the initial state in the subspace spanned by $\{\vert\uparrow_1\downarrow_2\rangle,\vert\downarrow_1\uparrow_2\rangle\}$, i.e. $\vert\psi(0)\rangle=(0,\cos{\xi},\sin{\xi},0)^{\text{T}}$ with $\xi\in[0,\pi/2]$. As shown in Fig. \ref{spin2}(a), the sum of $K_2^{\alpha}$ at the end of positive region is always less than $10^{-2}$ for various initial states in both cases (i) and (ii), we may neglect the contributions of the trajectories those underwent more than one normal quantum jump in calculating the averaged density matrix \cite{couplingcurves}.

In order to check the dissipative effect on the coherence of the system, we investigate the time-evolution of the entanglement between the spins. Here we employ the concurrence as the measure of entanglement which is given by $E_\rho=\max{\{0,\sqrt{\lambda_1}-\sqrt{\lambda_2}-\sqrt{\lambda_3}-\sqrt{\lambda_4}\}}$ with $\lambda_1\ge\lambda_2\ge\lambda_3\ge\lambda_4$ being the eigenvalues of the matrix $\rho(\hat{\sigma}^y\otimes\hat{\sigma}^y)\rho^{*}(\hat{\sigma}^y\otimes\hat{\sigma}^y)$ \cite{wootters1998}.

In Fig. \ref{spin2}(b) it is shown the time-evolution of the concurrence of the spins for three initial states $\vert\uparrow_1\downarrow_2\rangle$, $(\vert\uparrow_1\downarrow_2\rangle+\vert\downarrow_1\uparrow_2\rangle)/\sqrt{2}$ and $\vert\downarrow_1\uparrow_2\rangle$ in the subspace $\mathcal {H}_{\text{odd}}$. In the positive decay region, one can see that the entanglements demonstrates different behaviors depending on the initial configuration of the spins. This is due to the unbalanced dissipation on the spins. For initial separable states $\vert\uparrow_1\downarrow_2\rangle$ and $\vert\downarrow_1\uparrow_2\rangle$,
on the one hand the global operation generated by $\hat{H}_{\text{II}}$ will create entanglement between the spins, manifested by the increasing of $E_\rho$ at the beginning of the evolution. On the other hand the local dissipation on spin 1 tends to destroy the created entanglement via the normal quantum jump. Notice that the time-scale of the spin configuration of the system varies appreciably is $\sim \lambda_0^{-1}$ which is longer than $t_P$, so the initial spin configuration plays significant role in the early-stage evolution. As a consequence, the normal quantum jump is more likely to take place with the initial state $\vert\uparrow_1\downarrow_2\rangle$ since the probability of normal jump is proportional to the population of spin-up state. The entanglement created with initial state $\vert\downarrow_1\uparrow_2\rangle$ is robust against the local dissipation on spin 1.

From the solid lines in Fig. \ref{spin2}(b), one can observe a dip of entanglement around $t\approx 0.3$ showing the stronger dissipation with the initial $\vert\uparrow_1\downarrow_2\rangle$ configuration, also corresponding to the first peak of the temporal decay rate $\Delta(t)$ in Fig. \ref{decayrate}. By contrast, in the case of initial state $\vert\downarrow_1\uparrow_2\rangle$, the global operation dominates in the early stage of the time-evolution manifested by the monotonically increasing of the entanglement as shown by the dashed lines in Fig. \ref{spin2}(b). For the case of initial state $(\vert\uparrow_1\downarrow_2\rangle+\vert\downarrow_1\uparrow_2\rangle)/\sqrt{2}$, which is maximally entangled, the concurrence is monotonically decrease in the Markovian region since the effect of global operation is not significant within this time period.

When the decay rate becomes negative, the memory effect starts to impact via the reversed quantum jump. In the meantime the normal quantum jump is suspended. In the constant coupling case, under the actions of both the global operation and the reversed quantum jump, the coherence lost in the Markovian region is gradually recovered resulting in an obvious increasing of the entanglement as shown by the thin lines of Fig. \ref{spin2}(b). In the case of time-dependent coupling $\lambda(t)$, since the interaction between spins is switched off, thus the increasing entanglement is attributed to the reversed quantum jump although it acts locally, as shown by the thick lines in Fig. \ref{spin2}(b). This is a typical feature of non-Markovian dynamics that violates the divisibility of dynamical map \cite{rivasPRL2010}. In particular, for the case of initial state $(\vert\uparrow_1\downarrow_2\rangle+\vert\downarrow_1\uparrow_2\rangle)/\sqrt{2}$, the difference of the entanglement revival with and without spin interaction is negligible (of the magnitude $\sim10^{-3}$), as shown by the thin and thick dashed lines in Fig. \ref{spin2}(b), indicating the effects of global coherent spin-flipping operation on the symmetrically entangled state is slight.

In order to investigate the net contribution of the reversed quantum jump to the recover of entanglement, in Fig. \ref{spin2}(c) we show the time-evolution of entanglement with various initial state for $\lambda(t)$ in case (ii) which vanishes for $t>t_P$. In the Markovian region ($t\le t_P$), the asymmetry of the entanglement time-evolution about the initial spin configuration is obvious due to the unbalanced dissipation.
As entering into the non-Markovian region, the reversed quantum jump counteracts the effects of normal quantum jump leading to the recover of entanglement between spins. In particular, for $\xi/\pi\in(1/8,3/16))$ the sudden death of entanglement is observed as shown in Fig. \ref{spin2}(d). For $\xi/\pi=0.15$, the tail of the nonzero $\lambda(t)$ at the beginning of the negative decay region drives the entanglement into a sudden death of the entanglement \cite{yu2009}, however the memory effect recovers the entanglement through the reversed quantum jumps. This effect is only a consequence of the non-Markovian behavior of the open quantum system \cite{bellomo2007}.

\section{Comparison to the standard NMQJ method}
In the approaches of jumplike unraveling the time-local master equation, the observable is computed through the ensemble average over all the possible states of the system. Because the probabilities of the states are interdependent in the non-Markovian regime, as shown in Eq. (\ref{generic_traj}), computing these probabilities would be the most time-consuming part in the entire algorithm.

The standard NMQJ method is implemented by sampling a number of realizations with the identical initial (pure) state. In each realization, the system can undergo either the non-unitary evolution or normal/reversed quantum jumps. In order to approximate the true probabilities of the possible state, the number of realizations $N_r$ should be as large as possible. In the limit of $N_r\rightarrow\infty$, the existence probability in Eq. (\ref{ep_Kn}) is obtained. The improvement of our method lies in directly providing the existence probability. Moreover, one can choose an appropriate number $n^*$ to truncate the computation of the existence probabilities if the sum $\sum_\alpha{K_{n*}^\alpha}$ is below a threshold. For example, in Sec. \ref{sec_results}, we have truncated at $n^*=2$ for both models. In order to show the validity of the truncation in the modified method and its advantage in reducing the computational time, we compare $\sum_\alpha{K_2^\alpha}$, which is the total existence probability of the trajectory class $\{H_2^\alpha\}$, obtained by our modified method and the standard NMQJ method with various $N_r$.

\begin{figure*}[htpb]
	\centering
	\includegraphics[width=0.85\linewidth]{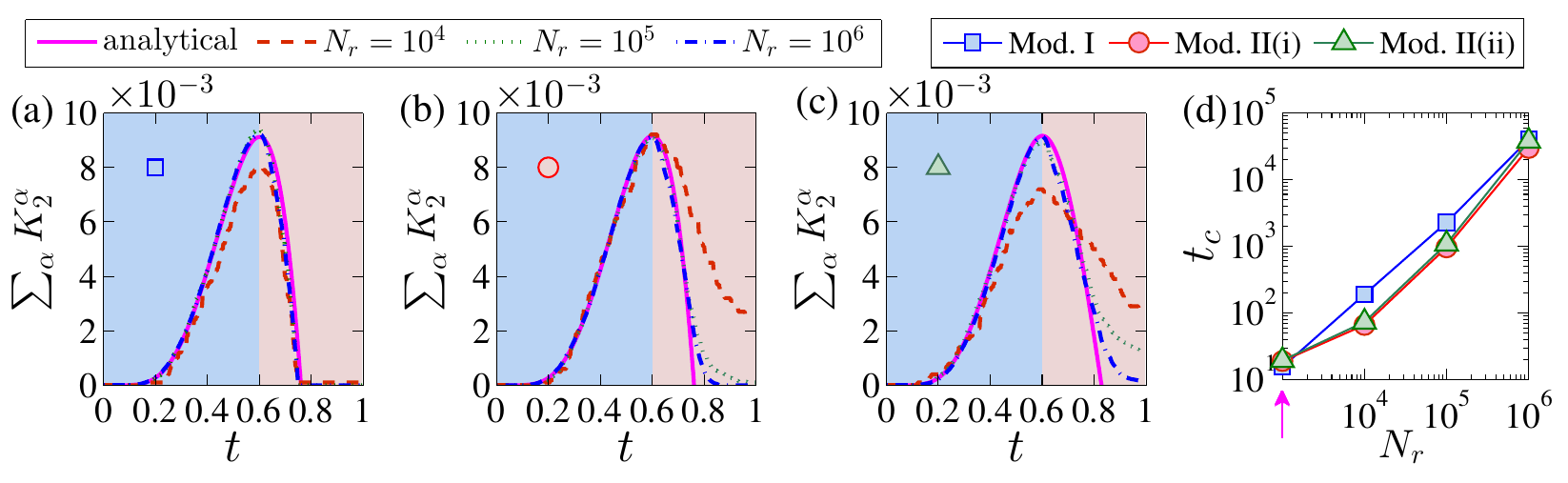}
	\caption{The total existence probability of the trajectory class $\{H_2^\alpha\}$ obtained by the modified NMQJ method and the standard NMQJ method with various realization number $N_r$ for (a) the Model I, (b) the case (i) of Model II, and (c) the case (ii) of Model II. The magenta solid line represents the result obtained from the modified method. The red dashed line, green dotted line, and blue dotted-dashed line represent for the results obtained from the standard method with $N_r=10^4$, $10^5$, and $10^6$, respectively. The consumed time $t_c$ (in the unit of seconds) of the simulation for each model are shown in panel (d). The blue square, red circle, and green triangle represents for the Model I, the case (i) and case (ii) of Model II, respectively. The horizontal axis of panel (d) denotes the realization number $N_r$. The consumed time of the modified method is plotted on the vertical axis of panel (d), indicated by the magenta arrow.}
	\label{compare}
\end{figure*}

In Figs. \ref{compare}(a)-(c), it is shown the time-evolution of the $\sum_\alpha{K_2^\alpha}$ for Model I, Model II in case (i) and case (ii), respectively. The continuous lines are computed according to the explicit expression while the discontinuous lines are computed by the standard NMQJ method with $N_r=10^4$, $10^5$ and $10^6$.
As shown in Fig. \ref{compare}(a), for the Model I the analytical result shows a good agreement with the results obtained by the standard NMQJ method. In particular, the value of $\sum_\alpha{K_2^\alpha}$ becomes zero at $t\approx0.72$ as predicted by the analytical expression, although the peak of $N_r=10^4$ case is a little lower than the analytical result. As shown in Fig. \ref{compare}(b), for the case (i) of Model II, the probability obtained with $N_r=10^4$ realizations deviates from the analytical result. As the $N_r$ increasing the probability converges to the analytical result gradually. Similar behavior can be observed for the case (ii) of Model II in Fig. \ref{compare}(c).

Now we investigate the time consumption of the modified and standard NMQJ methods. In Fig. \ref{compare}(d) the consumed time $t_c$ in the standard NMQJ method with various $N_r$ for both models are shown. The numerical computation is implemented on a regular computer workstation (Intel Xeon CPU E5-2680v3 and 64GB memory). As expected the time consumed in the computation grows dramatically as the $N_r$ increasing for the standard method. In order to obtain a more accurate result the $N_r$ should be larger than $10^5$ because there is still a non-zero tail at the end of the time-window for $N_r=10^4$. In this situation the $t_c$ is on the magnitudes of $10^3$ and $10^5$ seconds for $N_r=10^5$ and $10^6$, respectively, while the $t_c$ for the analytical result is on the magnitude $10\sim100$.

In order to estimate the accuracy of the modified NMQJ method, we compute the Bures metric of the final state computed by the modified and standard NMQJ methods. In the modified NMQJ method, the final density matrix $\tilde{\rho}$ is reconstructed according to Eq. (\ref{dm_construction}) with $n<2$. While in the standard NMQJ method, the final density matrix $\rho_m$ is obtained by averaging the states over all the $N_r$ realizations with $m=\log_{10}{N_r}$. The Bures metric of the state $\tilde{\rho}$ and $\rho_m$ is defined as follows,
\begin{equation}
D_B(\tilde{\rho},\rho_m)=\sqrt{2[1-F(\tilde{\rho},\rho_m)]},
\end{equation}
with $F(\tilde{\rho},\rho_m)=\text{tr}\sqrt{\sqrt{\tilde{\rho}}\rho_{m}\sqrt{\tilde{\rho}}}$ being the fidelity of two quantum states. The range of the Bures metric is $0\le D_B\le\sqrt{2}$. More specifically $D_B=0$ means two states are identical while $D_b=\sqrt{2}$ means two states are orthogonal. The Bures metric of $\tilde{\rho}$ and $\rho_m$ in Model I and both cases (i) and (ii) of Model II are listed in Tab. \ref{BuresMetric}. One can see that the final state computed by the modified NMQJ method (with truncation $n^*=2$) shows good agreement with the one computed by the standard method. In particular, as expected, the Bures Metric approaches zero gradually as the $N_r$ increases.

We finish this section with a brief discussion on the storage resources. In the implementation of our modified NMQJ method, the state of each trajectory is stored in a $D_H\times 1$ column vector where $D_H$ is the dimension of the Hilbert space ($D_H=2$ for Model and $D_H=4$ for Model II). At the end of the period of positive decay rate, $N_t$ state vectors are stored for recording the information of each trajectory. The $N_t$ is number of discretized time points in the period of $\Delta(t)>0$. For examples, the $N_t=601$ in the models of this paper. Thus the storage space can be evaluated as $D_H\times N_t$. In contrast, the storage space needed in the standard NMQJ method is evaluated as $D_H\times N_r$. Therefore, with almost the same accuracy, the modified method costs less storage resource.

\begin{table}
\centering
\caption{The Bures metric of the final state computed by the modified NMQJ method and by the standard NMQJ method with various realization numbers $N_{r}$. The parameters of each case are chosen the same as in Figs. \ref{compare}(a)-(c).}
\begin{tabular}{p{1.6cm}p{1.6cm}p{1.6cm}p{1.6cm}}
  \hline
  \hline
 $N_{r}$ & Model I & \multicolumn{2}{c}{Model II} \\
 ($10^m$)&&case (i)& case (ii) \\
  \hline
  $10^4$ & 0.0152 & 0.0491& 0.0499 \\
  $10^5$  & 0.0138 & 0.0104 & 0.0345 \\
  $10^6$  & 0.0162 & 0.0031 & 0.0130 \\
  \hline
  \hline
\end{tabular}
\label{BuresMetric}
\end{table}

\section{Summary}
\label{sec_summary}
In summary, we have proposed a modified NMQJ method to simulate the dynamics of open quantum system with the decay rate being either positive or negative. The proposed method unravels the time-local Lindblad master equation in stochastic quantum trajectories. By classifying all the possible quantum trajectories into the trajectory class, the memory effect featured by the negative decay rate can be implemented by the reversed quantum jump between a given trajectory class and its mother trajectory. We have derived the existence probability of each quantum trajectory which can be used for weighted sum in calculating the temporal expected value of the observable.

The existence probability directly gives the portion of the corresponding quantum trajectory in all the realizations in the limit of infinite $N$. Moreover the normalization of the existence probability is naturally satisfied by definition when all the trajectories are considered. It is shown that the existence probability can be calculated in a top-bottom manner, so in a realistic simulation one can only keep the trajectory class up to a truncation number $n^*$ provided that the contributions of those quantum trajectories undergo more than $n^*$ normal jumps are negligible to the sum. This makes our method more feasible from the perspective of costing less computational resources. The truncation can usually be adopted in the cases that, for examples, the driving field on the single-body system or the interaction between subsystem of a many-body system are not strong. Because the probability of normal quantum jump is proportional to the population of the upper level of the system and the amplitude of the decay rate. A normal quantum jump projects the system to the lower-level state, thus the population of upper level remains small and the next quantum jump is suppressed. Moreover, at the end of the period of positive decay rate, the amplitude of decay rate approaches to zero which further suppress the probability of normal quantum jump. So a quantum trajectory with multiply quantum jumps is unlikely to appear.

When the truncation is exact, that is there are no trajectories with more than $n^*$ jumps, our method coincides with the standard method. However, even if in this situation, our method still has advantages because it gives the explicit expression of the existence probability, while a sufficient large number of realizations are needed in the standard quantum jump method. On the other hand, if the existence probability of the trajectory with more than $n^*$ jumps is small but finite, the ensemble average can no longer fully reproduce the density matrix in the Lindblad master equation because part of the information is omitted due to the truncation. Here, we would like to emphasize that the efficient of the modified NMQJ for a more general system is restricted. For instance, in the quantum transport scenario where the quantum system interacting with two unbalanced baths, it is difficult to keep the history of the system because there will be an enormous number of quantum trajectories due to many quantum jumps may occur.

We have applied our method to two models of spin-1/2 system subject to the Lorentzian reservoirs. In both models the state after a normal quantum jump is not the eigenstate of the effective non-Hermitian Hamiltonian. The number of quantum trajectories is thus grow exponentially because the system can be excited by the external driving or the global operation which makes the conventional NMQJ method demanding in the numerical simulation. However, we found almost the weighted sum is contributed by the no-jump and one-jump quantum trajectories with the presented method. The revival of coherence of single spin and the entanglement of two spins are observed in the non-Markovian region. In particular the memory effects can recover the sudden death of entanglement through the reversed quantum jump.

We would like to note that the infinite-long memory time is assumed in deriving the existence probability. The dynamics with a finite-long memory time $\tau$ can also be investigated by modifying the probability of a reversed quantum jump, for instance, in Eq.  (\ref{233}) the denominator should be corrected as $\sum_{\alpha}{N_1^\alpha(t_P)}\rightarrow\sum_{\alpha}{N_1^\alpha(t_P)}-\sum_{\alpha'}{N_1^{\alpha'}(t_P-\tau)}$ where the element in $\alpha'$ is before the moment $(t_P-\tau)$.

In future work, it will be interesting to investigate the effect of non-Markovianity on the properties of quantum many-body systems, such as the information scrambling \cite{li_PRA_2020} and steady state of spin chains with boundary dissipation \cite{popkov_2020}. Finally, we hope the presented method may provide a potential way for efficiently simulating the non-Markovian dynamics of open quantum system.

\section*{ACKNOWLEDGMENTS}
We thank Dr. Giuliano Chiriac{\`o} for helpful discussion. J.J. would like to acknowledge support from the ICTP through the Associates Programme (2018-2023). This work is supported by National Natural Science Foundation of China under Grant No. 11975064.


\begin{thebibliography}{99}
\bibitem{couplingcurves}
The curves of the total existence probability $\sum_\alpha{K_2^\alpha}(t_P)$ as functions of $\xi$ for the cases of constant and time-dependent couplings almost overlap, because for $t<t_P$ the $\lambda(t)\approx\lambda_0$ as shown in the inset of Fig. \ref{decayrate}.
\end{thebibliography}
\end{document}